# Design of Core-Shell Structured Magnetic Microwires with Desirable Properties for Multifunctional Applications

*Sida Jiang,[†,‡] Tatiana Eggers,[‡] Ongard Thiabgoh,[‡] Claire Albrecht,[‡] Jingshun Liu,[§] Huan Wang,[∥] Ze Li,[†] Dawei Xing,[†] Weidong Fei,[†] Wenbin Fang,[†,⊥] Jianfei Sun,*[†] and Manh-Huong Phan*[‡]*

[†]School of Materials Science and Engineering, Harbin Institute of Technology, Harbin 150001, P. R. China

[‡]Department of Physics, University of South Florida, Tampa, FL 33620, USA

[§]School of Materials Science and Engineering, Inner Mongolia University of Technology, Hohhot 010051, P. R. China

[∥]Institute for Composites Science and Innovation, School of Materials Science and Engineering, Zhejiang University, Hangzhou 310027, P. R. China

[⊥]School of Materials Science and Engineering, Harbin University of Science and Technology, Harbin 150080, P. R. China

**ABSTRACT:**

**Amorphous Co-rich microwires with excellent soft magnetic and mechanical properties produced by melt-extraction technique are emerging as a multifunctional material for a variety of applications ranging from ultrasensitive magnetic field sensors to structural health self-monitoring composites. There is a pressing need for enhancing these properties to make the microwires practical for integration into new technologies. Conventional heat treatments**

at temperature below crystallization may improve the magnetic softness of an as-quenched amorphous wire, but usually deteriorate the good mechanical characteristic of the wire due to crystallization. To overcome this, we propose a new approach that utilizes the advantages of a multi-step Joule current annealing method to design novel (nanocrystal, amorphous)/amorphous core/shell structures directly from as-quenched amorphous microwires. These results show that the density and size of nanocrystals in the core can be optimized by controlling the Joule current intensity, resulting in the large enhancement of soft magnetic and giant magneto-impedance properties, while the amorphous shell preserves the excellent mechanical strength of the microwire. This study also provides a new pathway for the design of novel core/shell structures directly from rapidly quenched amorphous magnetic materials that are currently exploited in high frequency transformers, sensing and cooling devices.

# INTRODUCTION

Rapidly quenched ferromagnetic microwires are being targeted as multifunctional elements in a variety of exciting applications ranging from ultrasensitive magnetic sensors,[1,2] magnetoelastic sensors,[3] MRAM,[4] magnetic biosensors,[5] real-time respiratory motion monitoring devices,[6] magnetic hyperthermia cancer treatment,[7] to self-monitoring smart composites.[8] In particular, Co-based amorphous microwires with excellent mechanical, magnetic, microwave and giant magneto-impedance (GMI) properties have much to offer these applications.[9] Qin *et al.* have demonstrated a new concept of incorporating Co-based microwires with optimized GMI effect into fiber-reinforced composite as means of sensing stress and external magnetic fields through changes in impedance, which potentially offers an alternative to optical fibers for self-monitoring composites.[10] The advantage of using these microwires that exhibit the best characteristics for sensing tend also to have strength, thus contributing to the structure integrity of the composites. The research opens up new opportunities to develop smart structural composites with self-monitoring properties for a wide range of engineering and industrial applications.[11] By integrating a Co-based microwire as an excellent microwave absorber with the Fiber Bragg Grating (FBG) technology, Colosimo *et al.* have created a new class of microwave energy sensor with greater sensitivity as compared to an Au-based sensor.[12,13] Since the sensor is physically small and minimally perturbs the field being measured, it can be deployed as a distributed sensor. Inspired by a human hair skin structure, Zhang *et al.* have recently developed a multifunctional tactile sensor consisting of a pair of Co-based microwire arrays, for prospective applications in robotics and sensing in harsh environments.[14] To fully explore these applications, however, the microwires are required to possess not only enhanced magnetic properties but also good mechanical responses. While conventional heat treatment at temperature below crystallization has been shown to significantly improve the magnetic softness of an as-quenched amorphous microwire, it usually deteriorates its good mechanical characteristic.[15,16] Therefore, there must be a balance struck between the soft, uniform magnetic behavior and robust mechanical property to make these microwires practical for integration into new technologies. This represents an important but challenging task.[17]

Considering the multifunctional applications of the newly designed core/shell structures in which the two or more desirable functionalities that unfavorably coexist in a single component material can be achieved,[18-21] we propose a new strategy using a novel multi-step Joule annealing technique developed recently by our group [22,23] for the design and fabrication of a novel core/shell (nanocrystal, amorphous)/amorphous magnetic microwire from a single melt-extracted amorphous microwire of nominal composition $Co_{68.15}Fe_{4.35}Si_{12.25}B_{13.25}Zr_2$. Scanning electron microscopy (SEM) images of fracture patterns after failure by tensile loading uncover a core/shell structure of the annealed microwires. The devitrification of the metallic glass microwire core is examined by high-resolution transmission electron microscopy (HRTEM) with each Joule annealing step. We show, through control of the dc current intensity, that the bi-phase nanocrystal/amorphous structure of the core can be tailored by tuning the size and density of nanocrystals embedded in an amorphous matrix, yielding the desirable magnetic and GMI properties. Since Joule annealing has a majority effect on the core part (the dc current flows mainly inside the core layer), the shell structure of the wire retains essentially amorphous, thus preserving the excellent mechanical response of the as-quenched amorphous microwire. Our study paves a new pathway for the design and fabrication of core/shell structured microwires with desirable properties that fulfill the requirements of the emerging technologies.

**MATERIAL DESIGN**

In the metallic glass community, methods similar to Joule annealing, known as flash annealing,[24,25] use ultra-fast and uniform Joule heating from an RLC circuit to design nanocrystal composite microstructures. Heating rapidly to past $T_g$ (glass transition temperature) bypasses crystallization, while the approximately uniform volumetric heating of flash annealing is responsible for distributing the beneficial properties homogenously throughout the bulk sample. However, the sample needs to be quenched immediately to retain the microstructure formed during flash annealing.[25] In this study, we sequentially apply a dc current at temperatures below, near, and above $T_g$ to a ~45 ± 1 μm-diameter $Co_{68.15}Fe_{4.35}Si_{12.25}B_{13.25}Zr_2$ microwire (Figure 11a) at ambient temperature and pressure. The microwire was allowed to cool to ambient temperature before the next step was applied. The large specific surface area of these microwires implies a large heat transfer in the radial direction, therefore non-uniform temperature and cooling rate distributions. By starting the multi-step annealing process with low current intensity, low temperature structural relaxation occurs, which is typically mediated by short-range atomic motion. This has the effect of reducing residual stresses trapped in by fabrication and possibly forming more quenched-in nuclei, while leaving the macroscopic glassy nature intact. As the temperature is increased successively, the existing nanocrystals grow in regions of high temperature ($T > T_g -$ 100 °C), while in the cooler areas the crystal growth is slower but more quenched-in nuclei are being created. The temperature and cooling rate distribution becomes more important as the current intensity is increased. In the shell region of the microwire, lower temperatures and higher cooling rates preserve the as-quenched structure, while near the core, high temperatures and slower cooling rates ensure crystallization proceeds. The cyclic nature of this pattern, coupled with ultra-fast heating and a radially-dependent temperature and cooling rate distributions, contribute to a tunable core-shell two phase microstructure induced by the simple multi-step Joule annealing method. A general illustration of the fabrication process of a core/shell structured magnetic wire is shown in Figure 1.

In order to correlate simulations with experimental critical temperatures, the microwires were measured by differential scanning calorimetry (DSC) in the as-quenched state and after the 140 mA Joule anneal step. The exothermic peak area of the annealed microwire is larger than the as-quenched microwire due to structural relaxation and enhancement of medium and long range atomic order as seen in Figure 2a.[26] While there is no obvious difference in $T_{x1,2}$, after the 140 mA annealing stage, there appears to be a new metastable phase (MS III) made accessible by the annealing step of 140 mA. Based on electron diffraction, this phase is ascribed to $Co_2B$. The maximum temperature of the microwire at each Joule annealing step was estimated by considering the conservation of energy and iteratively solving for temperature.[27,28] Figure 2b shows that the microwire can reach a steady state temperature in less than a second. However, by Joule annealing the microwires in air and considering their small size, there must be a thermal gradient within the wire. The temperature distribution during Joule annealing at 140 mA is modeled using ANSYS software in Figure 2c. It shows only a core region of the wire experiencing high temperatures and the outer shell experiencing relatively low ones when compared to typical crystallization temperaures of $Co_{68.15}Fe_{4.35}Si_{12.25}B_{13.25}Zr_2$ products.

**RESULT**

**Microstructural evaluation.** Bright field high-resolution transmission electron microscopy (HRTEM) and selected area electron diffraction (SAED) are visual tools for analysis of material morphology as it can provide high-resolution structural and visual information. In this study, HRTEM images have been used to uncover the bi-phase nanocrystal/amorphous structure of the core of the annealed microwires. The microwire samples were thinned using two ion beams on either side in order to thin the selected region enough to reach electron transparency. The ion beam thinning was conducted at low energy and small controllable incident angle in order to manipulate thermal effects.

In the as-quenched state, ultra-fine nanocrystals (~2 nm) are found embedded in an amorphous matrix as seen in the HRTEM image of Figure 3a. These nanocrystals are surrounded by distortion and edge dislocation regions as displayed in the ring filtered inverse fast Fourier transform (IFFT) Figure 3(1) and Figure 3(2). Below Figure 3(2) is the twin filtered IFFT image of area 2. The distortion and edge dislocation regions (red circles and yellow rectangles in regions 1-2) as seen in these subfigures are caused by quenched-in nuclei, which are generated by local differences in rate and type of heat energy exchange, as well as stress inhomogeneity during fabrication.[29,30] In the corresponding SAED pattern (Figure 3I), several blurry diffraction ring patterns suggest a combination of amorphous and polycrystalline phases.

Figure 3b shows the HRTEM image after the 100 mA annealing step. During the 100 mA annealing stage, the highest temperature reached, T ≈ 515 °C, is near the core region. Judging by the yellow boxed regions 3-6 of the figure, there are some 1.5~2 nm medium-range order atomic clusters established which have developed from quenched-in nuclei. Additionally, the inset Figure 3(4) and 3(6) (corresponding to area 4 and 6) highlight regions of nanocrystal grain growth to about 3 nm. The distortion and edge dislocation surrounding nanocrystal are still present indicated by short-to-middle range clusters (yellow boxes) appearing in the twin filtered IFFT of area 4 and 6 (below Figure 3(4) and 3(6)).

The HRTEM image in Figure 3c was observed after the 140 mA anneal step. Compared to the 100 mA step, the nanocrystals are larger in size and quantity. The ring filtered IFFT image of area 7 (Figure 3(7)) shows the average size of the nanocrystals has increased to ~5.5 nm. The twin filtered IFFT images of area 8 and 9 (Figure 3(8) and 3(9)) display similar phenomenon as before, except the degree of internal lattice distortion has decreased slightly and edge dislocation has almost disappeared. Based on the calculated interplanar spacing and contrast with PDF cards, we confirm and mark the nanocrystal phase on Figure 3III. From this figure, the following polycrystalline phases can be identified: CoFe phase is cubic with (110) texture, CoSi and $Co_2B$ are (210) and (211), which have very close interplanar spacing (1.989 Å, 1.983 Å). The appearance of a small fraction of $Co_2B$ phase is seen the outermost ring in Figure 3III, which corresponds to the new crystallization peak (MSIII) from the DSC curve (Figure 2a) of this annealing stage.

After 160 mA annealing step, as is apparent from Figure 3d and 3IV, the nanocrystal size grows to ~9 nm with a strong diffraction ring. On account of the crystal being fully developed, there is almost no lattice distortion or edge dislocation as seen in ring filtered image Figure 3(10) from area 10.

The average diameter and auto-correlation function (ACF) estimation [22,27] of crystal volume fraction demonstrates a high correlation, and the ACF increases quickly with each annealing stage as shown in Figure 3e.

**Mechanical fracture morphology and properties.** Mechanical fracture experiments were performed on the microwires after each annealing stage. The as-quenched tensile strength (Figure 4, ~3612 MPa) is higher than a classic CoFeSiB amorphous microwire (~3400 MPa).[29] Similar pheonmenon also occurs in the 100 mA annealing stage, where the excellent tensile strength of ~4001 MPa is seen in Figure 4. For the 140 mA stage, the tensile strength (~3721 MPa) is slightly higher than the as-quenched stage. After annealing at 160 mA, the tensile strength markedly decreases to ~2691 MPa, however the plasticity is enhanced significantly (~4.2%).

In the as quenched state, the mechanical fracture morphology of the wire displays a typical vein pattern and slanted mechanical fracture.[31] There are dimpled defect structures and molten droplets on the surface of the fracture as shown in red framed area 1 and 3 of Figure 5b and 5c.

Figure 5d-5f shows the mechancial fracture after the 100 mA annealing step. As is apparent from the figure, thicker multi-shear bands and a relatively featureless zone appear after mechanical

loading. While the number of multi-shear bands has reduced from the as-quenched wire, their thickness looks like a shell structure. The main crack nearly disappears and replaces by a refined alveolate vein pattern,[32] the quality of pinholes and dimple structure is obviously improved. It is generally believed that these multiple structures are generated by Taylor instability and void coalescence.[33]

Figure 5g-5i shows the fracture of the 140 mA current annealing stage, in which the fracture changes from slanted to flat and multi-shear bands also disappear. This section of fracture displays a more obviously core/shell-like structure and a clear circular boundary as seen in Figure 5h and 5i. Upon further magnification (red boxed area 6), the ~3 μm thick shell still shows a vein pattern characteristic of amorphous fracture. However, the core area displays a section of feather-like fracture pattern with many relatively bigger pinholes similar to the classic crystal material.

With annealing current increased to 160 mA, the fracture occurs at a Rayliegh wave region. Thus a bowl-like structure is evident in Figure 5j. There is a very thick and smooth shell on the edge of the fracture (Figure 5k) and it appears different from the featureless region of a typical amorphous fracture and more like the area 5 framed in red in Figure 5h. There are rounded molten droplets on the core fracture surface and evidence of bigger nanocrystal cluster fracture (Figure 5l).

To demonstrate the necessity of the step-based Joule annealing technique in order to produce the core-shell structure, the tensile fracture, SEM, and HRTEM was conducted after a single 40 min long Joule anneal treatment at a current intensity of 140 mA. As can be seen in Figure S1(a), there is a similar feathered fracture pattern as in Figure 5g,h, however there is no core/shell structure evident. In the HRTEM image of the single step annealed microwire, there is a large nanocrystal formation of about 20 nm wide as seen in Figure S1(b). This is much larger than any of the nanocrystals observed in the stepped Joule annealing procedure.

**Magnetic and giant magneto-impedance properties.**

*Bulk and surface magnetization*. The bulk static magnetization of several melt-extracted wires with 15 mg total weight and approximately 3 mm in length were measured by VSM at room temperature with the magnetic field along the wire axis (Figure 6a (I)). As shown in Figure 6a (II), the bulk coercivity of the wire remains the same after the 140 mA annealing step but increases from $H_C \approx 4$ Oe to $H_C \approx 8$ Oe after the 160 mA annealing step. The dc permeability and coercivity are highly sensitive to the microstructure of a material, and an increase in bulk coercivity indicates devitrification over the majority of the microwire. However, Figure 6b shows the local coercivity measured by longitudinal Kerr effect on the surface of a single 3 mm long microwire decreases slightly from $H_C \approx 10$ to 8 Oe after the 140 mA annealing step. Then, after the 160 mA step was applied, the coercivity of the microwire did not appreciably change. This points toward less thermal activity at the surface as a result of the multi-step Joule annealing procedure.

*Surface domain structure*. Figure 7a-7c presents the AFM, MFM, and MFM scanlines, respectively, of the same region of wire during each annealing step. In order to ensure the same area for observation after each annealing step, a surface feature located with AFM (shown in area 1 of Figure 7a) marked the tagged center point of the scans. After 100 mA current annealing, micron-sized atomic clusters appear on the surface as seen in area 2 of Figure 7a. In the 140 mA annealing stage, another cluster nucleates (area 3 of Figure 7a) while prior atomic clusters disappear. The roughness parameters and average leakage field level $M_a$ sharply reduces during this annealing stage as seen in Figure 8a and 8b. As the annealing current increased to 160 mA, larger atomic clusters (area 4 and 5 of Figure 7a) formed on the surface. Figure 7b shows MFM images of the surface after annealing *in situ*. Joule annealing can effectively reduce the defects in the magnetic domain structure as shown in area 6-8 of Figure 7b. At the same time, a connection between adjacent domains is seen forming in the corresponding area 9 and 10 of Figure 7b. This has the impact of increasing the surface roughness, decreasing the leakage field level and destroys the periodicity of the domain structure as shown in Figure 7c.

*Giant magneto-impedance*. The penetration depth of an ac current in a conductor like copper is around 65 µm at 1 MHz. However, in a ferromagnetic conductor with similar resistivity and a high permeability, the penetration depth of the current can drop by an order of magnitude to just a few microns. Therefore, by applying a small magnetic field to increase the permeability of the ferromagnetic conductor, the skin effect will increase significantly and the majority of the current will be distributed in a thin shell (Figure S3). This is the underlying principle of the giant magneto-impedance (GMI) effect. As an investigative tool, the magneto-impedance spectra in a frequency range of 1-110 MHz will mainly reflect the skin depth change due to the change in magnetic permeability of rotational magnetization processes, since most domain wall movement is damped in wires starting in the low MHz. One can correlate high permeability to the amorphous magnetic phase, while a decrease in rotational permeability would correspond to a crystalline magnetic phase.

The 3D images of magnetic field- and frequency-dependent GMI profiles for different annealing stages are shown in Figure S2 and Supplementary Note 1. The change in impedance relative to zero magnetic field, $\Delta Z/Z_0$, is plotted at the frequency in which the maximum change is reached, $f_K$, in Figure 9a for each annealing stage. The maximum change in impedance for the as-quenched wire is about 51% at $f_K$ = 25 MHz as indicated in Figure 9a. When the annealing stage reaches 100 mA or ~515 °C near the core, the maximum change shifts down to 19 MHz. With subsequent annealing stages, as the core temperature approaches ~691 °C (140 mA), the maximum value of $\Delta Z/Z_0$ reaches ~567%, which is nearly 10 times the as-quenched value. However, the value is found to reduce to ~351% after the 160 mA annealing step or a core temperature ~801 °C. The sensitivity of magnetic field, $\xi$, is shown in Figure 9b. Similar to the $\Delta Z/Z_0$, the sensitivity gradually increases to ~755 %/Oe after 120 and 140 mA annealing then decreases after the 160 mA annealing step. As is apparent from Figure 9c and 9d both the resistance and inductive reactance have a similar tendency as the impedance.

Figure 10a magnifies the low field region of the GMI curves presented in Figure 9a. What is shown here is the linear detection region increases with each annealing step until the 160 mA step. Figure 10b shows that the $H_K$ field, that is the field location of the maximum change in impedance, gradually increases in magnitude with annealing and is symmetric about zero field. The $f_K$ decreases also with annealing. In Figure 10c, the impedance spectra is presented at $H_{ex}$ = 0 Oe and $H_{ex}=H_K$, at the $H_K$ fields indicated in Figure 10b. The most important feature of this plot is the distinct flattening of the impedance at zero magnetic field, while the maximum impedance is more modestly modified by the annealing current.

In order to correlate the GMI with the skin depth and understand how the annealing treatment modifies the penetration depth of the excitation current, the skin depth is calculated based on a simple geometrical model using the dc and ac resistance of the wires after each annealing step.[34] The analyzed results are shown in Figure S3 and Supplementary Note 2.

**DISCUSSION**

As is apparent from Figure 3I, the SAED pattern of the as-quenched wire shows that the formation of nanocrystals is eutectic crystallization during the solidification process.[30] For this phenomenon as mentioned different rate and type of heat energy exchange and stress inhomogeneity during fabrication. The another reason is high melting point elemental (Zr) doping will enhance the Gibbs free energy of the liquid alloy, further increasing the critical radius $r_c$ of quenched-in nucleus by classical nucleation theory. Some short-range and few medium-range order atomic clusters are seen in the as-quenched wire. However, considering the fracture morphology of the wire, along with shear banding at the surface,[31] we conclude that the as-quenched wire is mainly an amorphous phase with small, ~2 nm nanocrystals that precipitate the near surface area and distribution quenched-in nuclei. In fact, from SEM images of the fracture, it can be seen that the main crack is refined and impeded near the edge of wire (Figure 5b area 2 and 3), which close related to mechanical strength. It also further indicates a high concentration of as-quenched nuclei and small nanocrystals near the surface of the wire as a result of the fabrication process. In the highlighted areas 3 and 4 of Figure 5b and 5f, the pinholes and dimple structure are shown to concomitantly exist on the alveolate vein-pattern. It is generally believed that pinhole structures are caused by the development of nanoscale pinholes which generated by the accumulation of free volume and defects of flowing deformation.[35] In the present research, the as-quenched nuclei and small nanocrystals will promote the development of this situation, and impede main crack expansion then enhance the mechanical strength. Therefore, the as-quenched tensile strength of the melt-extracted mircowire in this work is higher than the traditional CoFeSiB amorphous microwire.

At the 100 mA stage, the highest temperature is near ~515 °C, which is below $T_g$ and $T_x$ of any crystalline products seen by DSC. At this temperature ($T_g - 100$ °C~$T_g$), high temperature structural relaxation occurs through relatively long-distance diffusion of defects in the amorphous structure to improve chemical short-range order, while leaving the macroscopic glassy structure intact.[36] Nonthermal activated nucleation is by nature heterogeneous nucleation, whose dynamics is closely related to fabrication technique, thermal history, and the quantity of subcritical as-quenched nuclei. Many existing atomic clusters such as the subcritical dimensions and undeveloped as-quenched nuclei (~1 nm) will develop into normal as-quenched nuclei; normal as-quenched nuclei began grown through fusing surrounding atoms and nuclei forming bigger nanocrystal and medium-range order atomic cluster. In comparing the IFFT reconstruction of area 1 & 3, 5 of Figure 3a and 3b, one can see similar symmetrical patterns of one or two atoms and holes centered in the as-quenched and 100 mA annealed wires. Such a symmetrical structure is also found in Ref. [37], there is no doubt that these medium range-order structures correspond to the sphere, and two fused spheres in three-dimensional space. Based on the proposed quasi-equivalent closely packed model,[38] the 3D shape of this structure is icosahedron or similar to it. Low temperature structural relaxation of 80 mA annealing ($T_g$ – 200 °C~$T_g$ – 100 °C) can be considered a preprocessing stage with topological short-range, local atomic motion that removes some residual stress.

As shown in areas 4 and 6 of Figure 3b, the nanocrystal grain size has grown to ~3 nm, and there is still some as-quenched nuclei accompanied around its edges to cause lattice distortion and edge dislocation to appear on nanocrystal edge (below Figure 3(4) and 3(6) yellow box). It would have a big influence on mechanical properties. Since the size of nanocrystals in this work are smaller than 10 nm, the stability of the grain boundary is more suitable than the Hall-petch relationship to explain the hardening and excellent tensile strength ~4001 MPa of 100 mA.[39] Such a hardening can be attributed to grain boundary stabilization through relaxation and segregation of the nanocrystals.[39] The high concentration of distortion and edge dislocation are generated by the as-quenched nuclei around poorly developed nanocrystal boundaries. Such a nanostructure will refine the main cracks and prevent their tendency to form large alveolate vein patterns. The initiationd and evelopment of multi-shear bands tend to extend to the wire inside due to the shear slip and void coalescence in the wire edge region. The increase of nanocrystal size in the core and increase in ACF fraction show this. So the shell-like structure is observed, while the number of multi-shear bands is reduced. The size of the crack pattern is closely related to strength based on Griffith theory, as discussed in Ref. [40]:

$$\gamma_p = \frac{1}{6\pi}\left(\frac{k_c}{\sigma_y}\right)^2 \qquad [1]$$

$\gamma_p$ is size of crack pattern, $\sigma_y$ is the tensile strength, they present obvious inverse proportion relationship. The surface roughness also does not show much change after the 100 mA anneal step, indicating rarely crystal growth on the surface, while the inner ACF fraction enhances from 12.5% (as-quenched) to 20.3% as shown in Figure 3e. Although there is not much change in the surface magnetic properties after 100 mA annealing, but the magneto-impedance spectra (and $\Delta Z/Z_0$) largely influenced as seen by a drop in the impedance at zero magnetic field. This indicates a tendency for increase in penetration depth of the exciting ac current, which could be due to a decrease in resistivity (there is some crystallization evident by ACF) and increase in rotational permeability through atomic scale movement of magnetic atoms during structural relaxation.[23] Therefore, the average level of leakage magnetic field ($M_a$) shows a dramatic decline, and the roughness parameters show a slight decrease.

With increasing the annealing current to 140 mA, the maximum temperature in the core is near ~691 °C, which is past $T_g$ and $T_{x1,2}$. Both nanocrystals and as-quenched nuclei in this temperature region begin growth through fusing with surrounding atoms and nuclei to form larger nanocrystals (~5 nm) and medium-range order atomic clusters. Consequently, the nanocrystal ACF fraction increases to nearly 37.5%. At this annealing stage, the nanocrystal and amorphous core/amorphous shell phase separation becomes most evident in the fracture images, which suggests that some sort of cooperative diffusion has taken place. With the core experiencing temperatures around $T_g$, the defects can combine or annihilate each other, while near the surface of the wire the temperatures may be low enough not to change the major population of defects. When temperatures reach $T_g$, homogeneous (thermal-activated) nucleation occurs concomitantly with some inhomogeneous nucleation and eutectic and polymorphous crystallization. Moreover, based on the phonon scattering classical theory, with increases in nuclei and nanocrystal regional resistance declines, which results in more current flowing through the core area. Continuing the tendency of the previous stage, the impedance at zero magnetic field dramatically reduces brought about by the large increase of the skin depth, which enhances the GMI ratio.

Accordingly, in case the large supercooling degree and crystallization temperature much lower than the melting point, the rate of nucleation and growth can be given by the Arrhenius equation:

$$I \approx I_0 \exp\left(-\frac{Q_n}{RT}\right) \qquad [2]$$

$$U \approx U_0 \exp\left(-\frac{Q_g}{RT}\right) \qquad [3]$$

thermal activation energy ($Q$), supplied by the dc current, plays a key role in the average diameter and ACF fraction determined by HRTEM. Due to internal heat transfer differences in wire, there will be some diversity in nanocrystal diameter. In crystal dynamics view, CoFe phase growth excludes Si and B atoms, CoSi and $Co_2B$ growth excludes Fe atoms; there must have interactive diffusion in the front of growth interface, therefore exclusive atoms move to the front of corresponding phase for providing growth conditions.

Accordingly, the fracture morphology of the core region has changed significantly from the previous stages, which change from slant to flat and multi-shear bands is also disappear, revealing intergranular fracture and soft edges induced by localized melting. During fracture, nanocrystals attract free volume and other defects to form many tiny crack sources. Furthermore, the stored inner strain energy will promote the growth of nanocrystals and fuse adjacent nanocrystals (Figure 5h, area 5). However, there is still some lattice distortion remaining near the nanocrystal boundary based on the HRTEM analysis, so the tensile strength (~3721 MPa) is slightly higher than the as-quenched value (~3612 MPa). Around the edge of the wire there is evidence of vein patterns typical to amorphous fracture and some pinhole defects. The fracture morphology, along with the skin depth increase from ~3 μm of as-quenched (~5 μm, 100 mA stage) to ~12 μm at $H_{DC}= 0$, provide evidence of a (nanocrystal, amorphous) core /amorphous shell structure. The surface atomic dynamic of an amorphous wire is very reactive under the action of thermal and magnetic field generated by an annealing current.[41] Therefore, surface magnetic atoms have a more uniform distribution, as seen by the roughness parameters and $M_a$ decreasing to the lowest level. The surface domain is more even as shown in the MFM scanlines (Figure 7c, 140 mA stage).

At the highest annealing current of 160 mA, $T_{max} \approx 800$ °C, continued crystal growth is seen and the average crystallite size of ~9 nm provides a strong diffraction ring and ACF fraction enhancement to over 50 %. Thermal energy exceeds all metastable phases (MS I, II, III) to reach a stable phase, as related by the DSC curve. After tensile fracture, the presumably melted globs near the core have increased in size, which correlates with the HRTEM estimation of crystallite size. The smooth and featureless fracture around the edge of the wire has increased in thickness. At the same time, the fracture has occurred on a thick portion of the wire, known as a Rayleigh wave formed during fabrication. In as-quenched microwire the fracture usually occurred on a thin area just like the below area of this bowl structure.[42] This location of tensile failure is likely due to the large sized nanocrystals that accumulate in this region. The induced plasticity could be for two reasons. One is macroplasticity caused by the plastic shear on a realitively larger region and the other is intercrystalline rupture appears to bigger size nanocrystal on raised force area (bowl-like structure). The thick shell ruptures first and then the force area gradually reduces, the tensile curve displays deviates from the linear on a very short strain. Meanwhile, the bigger atom clusters formed on surface, which increase the coercivity and surface roughness, $M_a$. Large clusters can annex or divide adjacent domains in some more surface dynamic areas, breaking the periodicity, which results in the impediment of domain rotation under external magnetic field. So, the skin depth at zero field (Figure S3a) also reduces then decreasing the GMI ratio.

**CONCLUSION**

We have shown a new approach that utilizes the advantages of a multi-step Joule current annealing method to design novel (nanocrystal, amorphous)/amorphous core/shell structured magnetic microwires from melt-extracted Co-rich amorphous microwires. SEM and HRTEM studies have confirmed the formation of the core/shell structure of a melt-extracted $Co_{68.15}Fe_{4.35}Si_{12.25}B_{13.25}Zr_2$ microwire as a test sample. The HRTEM analysis has also shown that the density and size of nanocrystals in the core can be tuned and optimized by varying the dc current intensity. Relative to the as-quenched microwire, the optimal core/shell structured microwire (e.g. the 140 mA annealed microwire) possesses enhanced mechanical, magnetic and GMI properties. These enhanced properties make the microwire a novel and multifunctional component that fulfils the requirements of the emerging technologies [45-48]. Our proposed approach can also be extended to designing novel core/shell structures of other types of rapidly quenched magnetic materials.

**EXPERIMENTAL SECTION**

The high quality as-quenched $Co_{68.15}Fe_{4.35}Si_{12.25}B_{13.25}Zr_2$ amorphous wires possess the diameters ranging from 20 to 60 µm in 20~50 cm length were fabricated by melt-extraction described elsewhere,[22,42] and saturated magnetostriction with nearly a small negative value (nearly $|\lambda_s| \leq 10^{-6}$).[29]

The uniform and continuous wire with diameter of ~45 ± 1 µm and ~26 mm in length (Figure 11a) was selected and post-processed by multi-step DC Joule annealing with current amplitudes ranging from 80 mA (current density ~$5.03 \times 10^7$ A/m$^2$) to 160 mA (~$10.07 \times 10^7$ A/m$^2$), step value was set as 20 mA and annealing time is 10 min for each stage on this single wire [7]. The impedance of wire was measured on an Agilent 4294A precision impedance analyzer in room temperature, which frequencies of the driving current were varied from 1~110 MHz and amplitude of the driving current was kept at 20 mA. The samples were placed in a pair of Helmholtz coils parallel to the axis of the microwire but transverse to the geomagnetic field. The GMI ratio, $\Delta Z/Z_0$, is defined by the same way as reported in Ref. [43]:

$$\frac{\Delta Z}{Z_0}(\%) = \left[\frac{Z(H_{ex}) - Z(H_0)}{Z(H_0)}\right] \times 100\% \qquad [4]$$

and field sensitivity of GMI ($\xi$) can be expressed as the following expression:

$$\xi(\%/Oe) = \frac{d[\Delta Z/Z_0]}{dH_{ex}} \qquad [5]$$

where $Z(H)$ is the impedance at external field $H_{ex}$. The impedance ratio is normalized by $Z(H_{ex}=0)$. The $H_{ex}$ location of the maximum change in impedance is denoted by $H_K$, which is typically referred to as the anisotropy field.

The magnetic force microscopy in this work was conducted on a Bruker Dimension Icon in tapping-lift mode with the cantilever driven slightly below its natural resonance frequency to maximize the change in oscillation amplitude. The basic theory and the details of the processing procedure for MFM images are shown in Ref. [44].

The thermal properties associated with glass transition and crystallization were evaluated by differential scanning calorimetry (DSC; TGA/SDTA85IE) under a flowing argon atmosphere at a heating rate of 20 °C min$^{-1}$. The measurement of magnetic properties was carried out on a vibrating sample magnetometer (VSM; Lake Shore 7410) and homemade Kerr Effect magnetometer in the longitudinal geometry at room temperature. The average temperatures of wires during annealing can be evaluated using the energy conservation law and value iteration as follows:[27,28]

$$\rho_M c \frac{dT}{dt} - Q_E - Q_h - Q_\Gamma = 0 \qquad [6]$$

where $\rho_M$ is the microwire density, $c$ is specific heat capacity, $Q_E$ is DC thermal efficiency, $Q_h$ and $Q_\Gamma$ are the microwire surface convective and radiant heat exchange amount. Simulation data from Fortran programming and Ansys software was used to simulate the thermostatic temperature field distribution of wires.

The structural information of wires was examined by scanning electron microscope (SEM; Hitachi S-4700). In addition, the nanostructure of as-quenched and annealed wires was examined by high-resolution transmission electron microscopy (HRTEM; Tecnai G2F30) and the steps for sample preparation are shown in Ref. [27]. HRTEM images of the prepared wire sample are shown in Figure 11b. Digital-micrograph software was used to analyze the HRTEM data. Tensile samples were prepared conforming to the ASTM standard D3379-75 with a gauge length of 10 mm and diameter range from 20~50 μm with the annealing current density constant. Micro-tension experiments of wires with different area reductions were performed using a 10 N Instron 3343 universal testing machine at a constant strain rate of 0.252 mm min$^{-1}$.

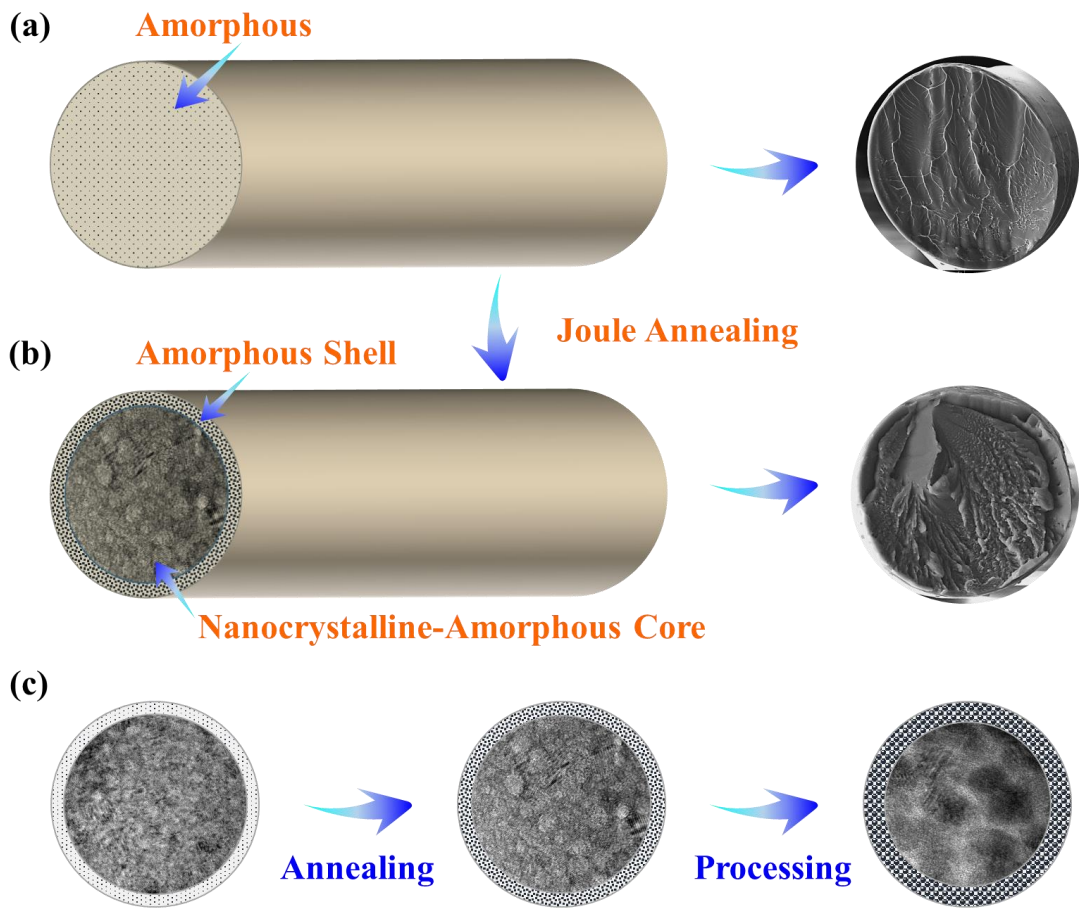

**Figure 1.** (a) A schematic of the amorphous as-quenched microwire along with the cross-sectional SEM image. (b) A schematic of the microwire after multi-step Joule annealing. The core area capture form HRTEM. To the right is a cross-sectional SEM image. (c) Schematic of the core-shell structure section appearing after multiple annealing steps.

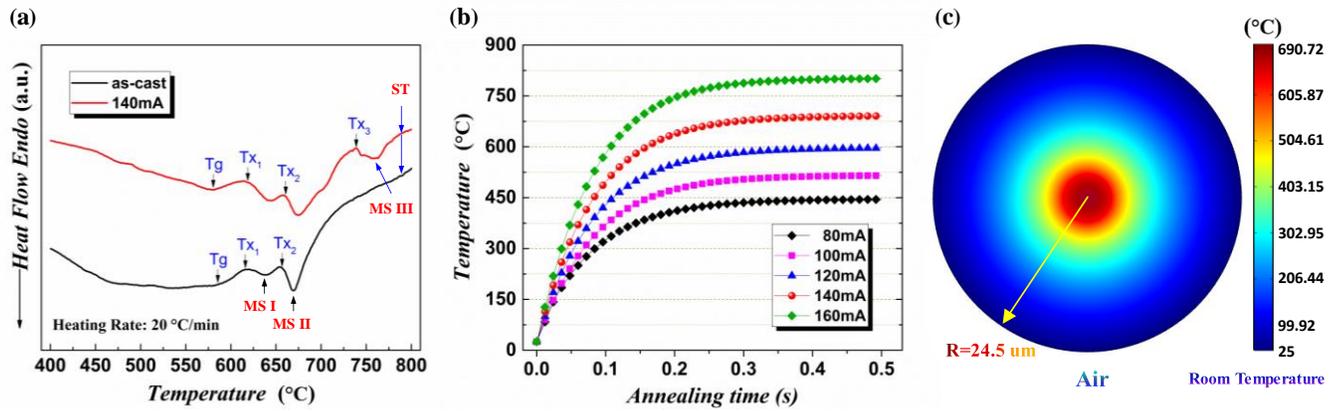

**Figure 2.** (a) DSC curve of the as-quenched microwire (black line) and 140 mA current annealing stage (red line), $T_x$ is crystallization temperature, MS is metastable phases and ST is stable phase. (b) Simulation of the core temperature as a function of time for each annealing stage. (c) Simulation of the temperature distribution in the microwire during 140 mA current annealing stage.

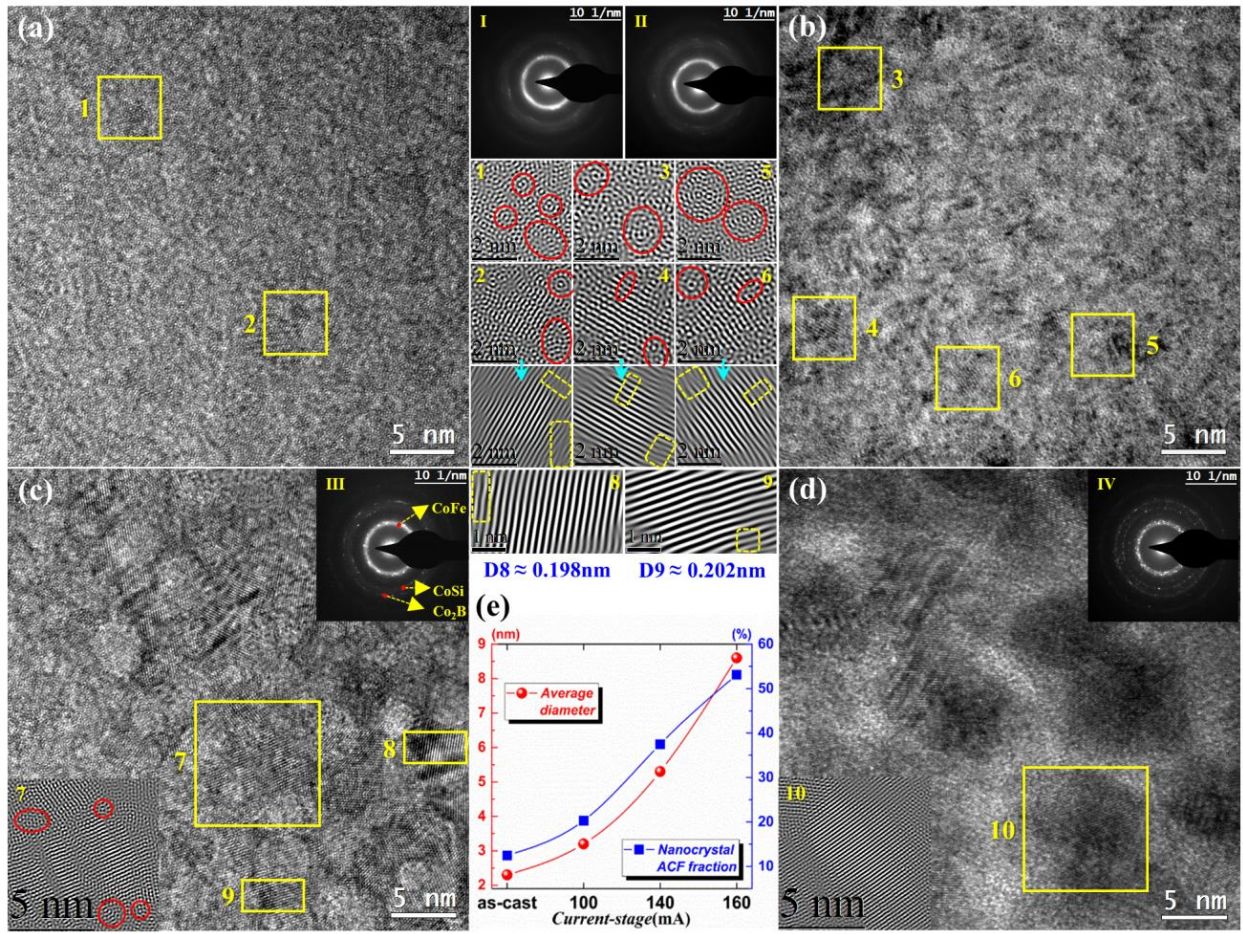

**Figure 3.** HRTEM images and SAED of the microwire and the corresponding local magnifications: (a), (b), (c), and (d) show the HRTEM image of the as-quenched, 100 mA, 140 mA, and 160 mA annealing stages, respectively. Roman numerals I-IV display the corresponding SAED image. Insets 1-10 show the ring- and twin- filtered IFFT images of local magnification area framed in yellow box. The unlabeled insets below 2, 4, and 6 show a twin filtered IFFT image of the same regions 2, 4, and 6, respectively. Insets D8 and D9 show an estimation of the interplanar distance of the nanocrystal structures observed from insets 8 and 9. (e) shows the results of the ACF calculation along with the average nanocrystal size as a function of current stage.

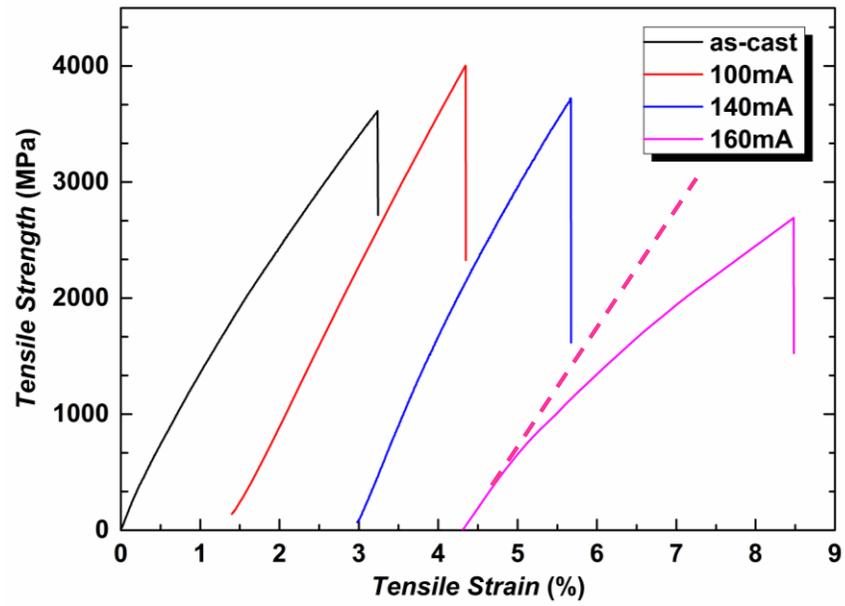

**Figure 4.** Tensile stress-strain curve for the microwire in the as-quenched state and after 100 mA, 140 mA, and 160 mA annealing stages.

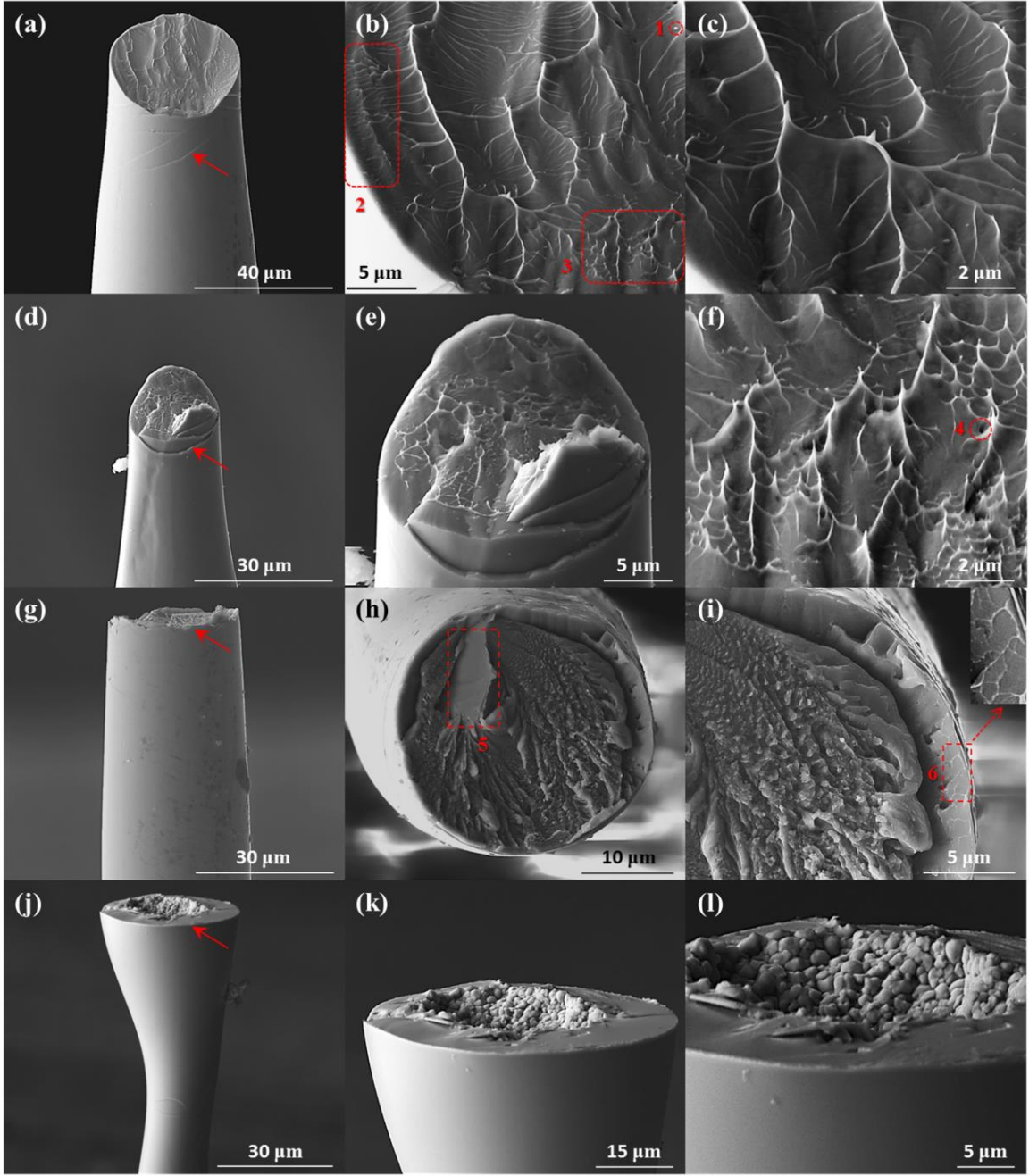

**Figure 5.** Mechanical fracture SEM images with different magnifications for (a)-(c) as-quenched microwire, (d)-(f) 100 mA annealing stage, (g)-(i) 140 mA annealing stage, and (j)-(l) 160 mA annealing stage. Location of characteristic region insets are marked in red box.

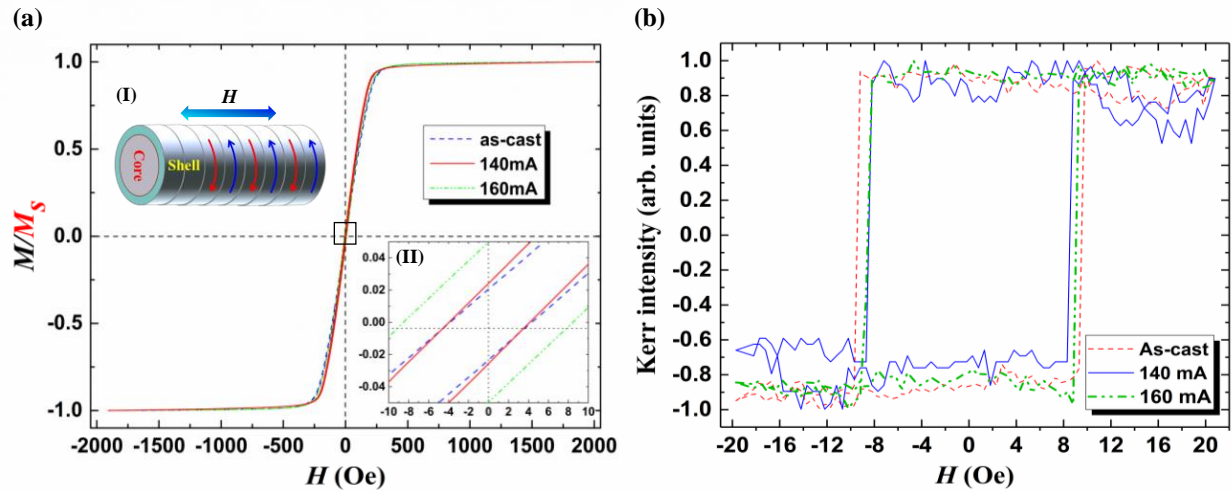

**Figure 6.** (a) Normalized magnetization curves from VSM for the as-quenched (blue dash), 140 mA (red), 160 mA (green dash) microwires. The inset (I) shows the direction of external magnetic and (II) shows the coercivity at low field. (b) Normalized magneto-optical Kerr effect hysteresis loops in the longitudinal geometry of the microwire surface for the as-quenched wire (red dash), 140 mA-treated (solid blue) and 160 mA-treated (green dash).

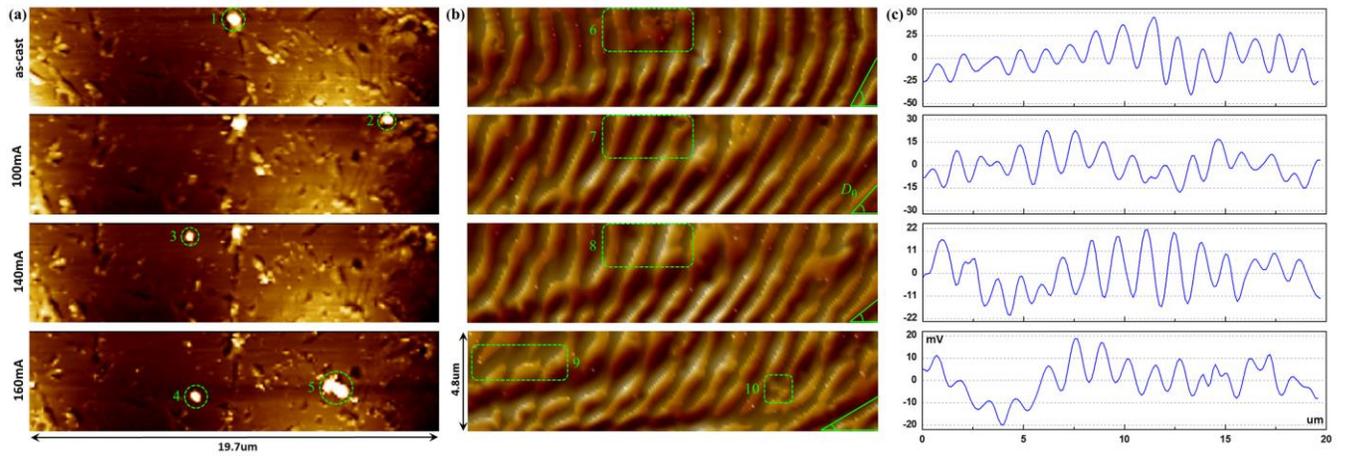

**Figure 7.** Observation of (a) surface atomic structure by AFM, (b) magnetic surface structure by MFM and (c) MFM period scanlines of the as-quenched wire, 100 mA-treated, 140 mA-treated, and 160 mA-treated microwires, from top to bottom, respectively. Feature points 1-10 discussed in the text are marked in green.

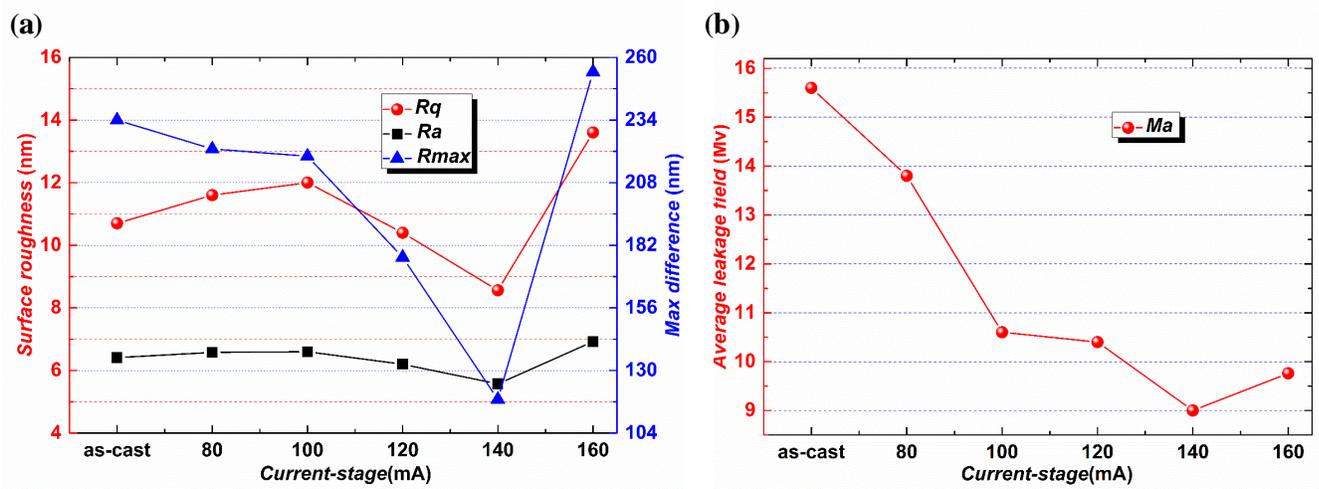

**Figure 8.** Quantitative parameters related to the AFM and MFM images. (a) arithmetic mean roughness $R_a$, root mean square roughness, $R_q$, and maximum roughness height $R_{max}$; (b) average strength of leakage magnetic field $M_a$, and angle between domain and axial direction $D_\theta$ for each annealing stage.

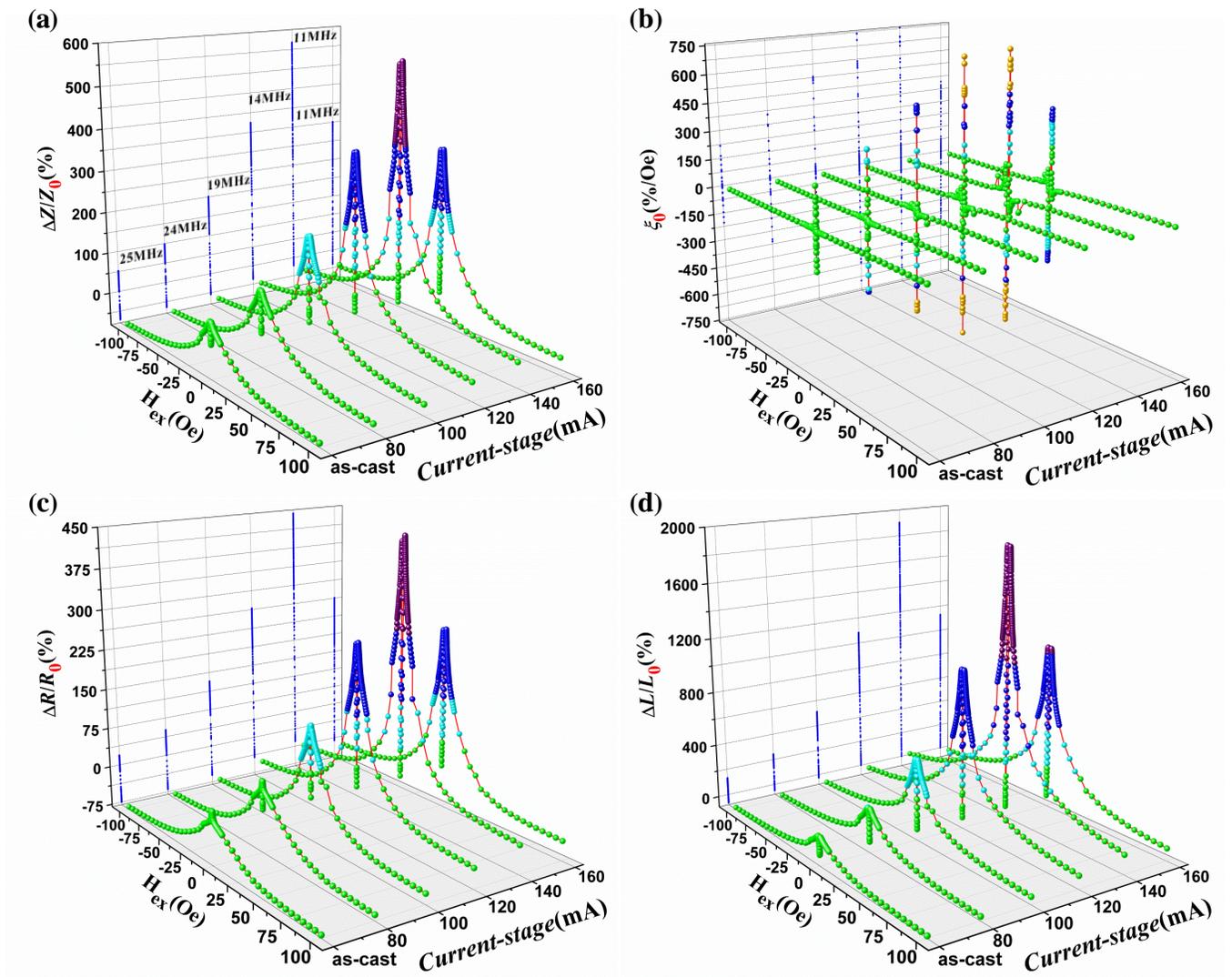

**Figure 9.** (a) The change of impedance ratio relative to zero magnetic field for the microwires after each current annealing stage and the applied external DC magnetic field $H_{ex}$. The corresponding frequency of the measurement is listed in the figure. (b) The sensitivity of the microwire after each annealing stage. The change in resistance (c) and change in reactance (d) relative to zero magnetic field.

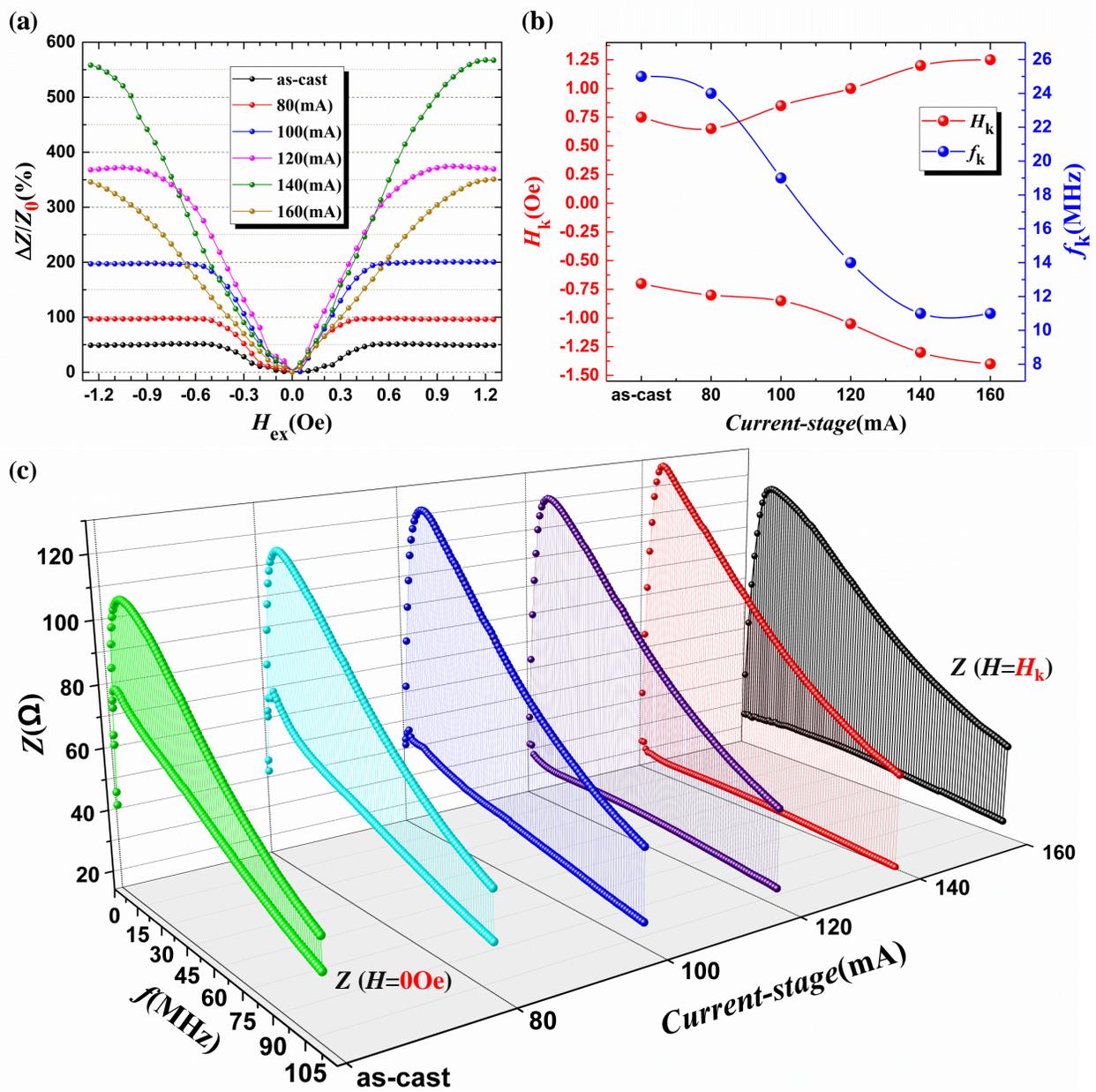

**Figure 10.** (a) The low-field, linear region of the change in impedance shown in the previous figure. (b) The value of the effective anisotropy field, $H_K$, and the excitation frequency where the impedance ratio was a maximum. (c) The frequency dependence of the impedance at $H_{ex} = 0$ Oe and $H_{ex} = H_K$ for each annealing stage.

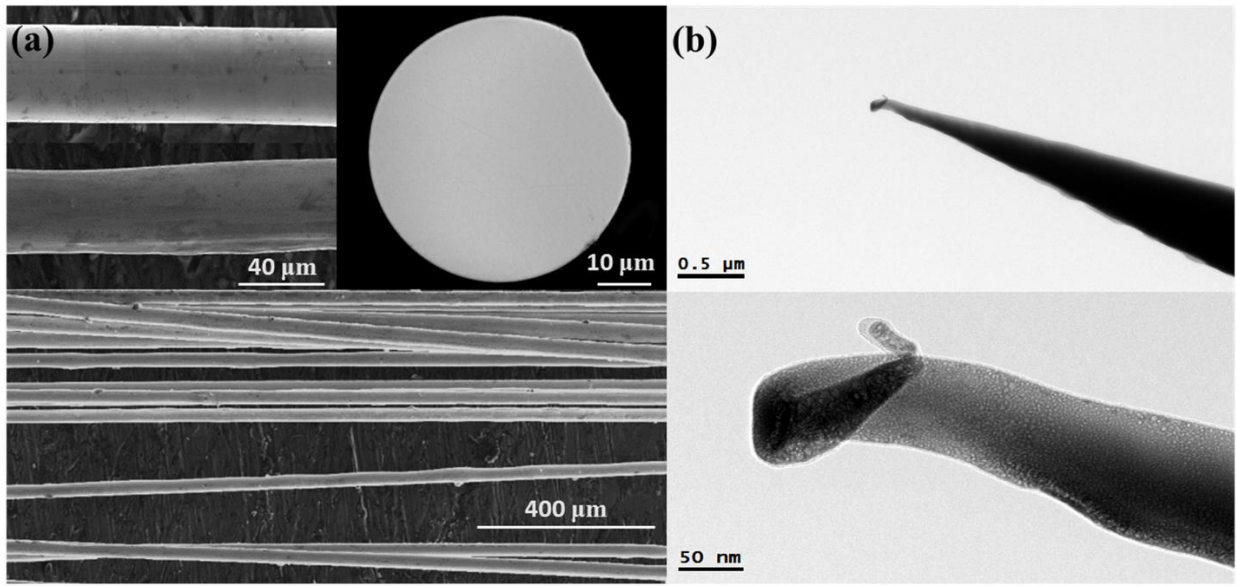

**Figure 11.** (a) SEM images of the melt-extracted microwire samples at different magnifications. (b) TEM images of the microwire after ion beam thinning to show the region measured by HRTEM.

## ASSOCIATED CONTENT

**Supporting Information**

3D images of GMI properties, skin depth calculate images and corresponding analysis.

## AUTHOR INFORMATION

**Corresponding Author**

*E-mail: phanm@usf.edu (M.H.P), *E-mail: jfsun_hit@263.net (J.F.S).

**Author Contributions**

The manuscript was written through contributions of all authors. All authors have given approval to the final version of the manuscript. S.D.J. and T.E. contributed equally to the work.

**Notes**

The authors declare no competing financial interest.

## ACKNOWLEDGMENT

This work was financially supported by the National Natural Science Foundation of China (NSFC) under grant Nos. 51371067 and 51671071. S.D.J. acknowledges support from the China Scholarship Council (CSC) fellowship. J.S.L acknowledges National Natural Science Foundation of China (NSFC) under grant No. 51561026. Research at USF was supported by the U.S. Department of Energy, Office of Basic Energy Sciences, Division of Materials Sciences and Engineering under Award No. DE-FG02-07ER46438 (MOKE and GMI measurements).

**Table of Contents:** A new approach has been developed that utilizing a multi-step Joule current annealing method for the design and fabrication of novel (nanocrystal, amorphous)/amorphous core/shell structures directly from as-quenched amorphous magnetic microwires. These core/shell structured microwires possess enhanced mechanical, soft magnetic and giant magneto-impedance properties that fulfil the requirements of the emerging sensor technologies.

**Abstract Graphic**

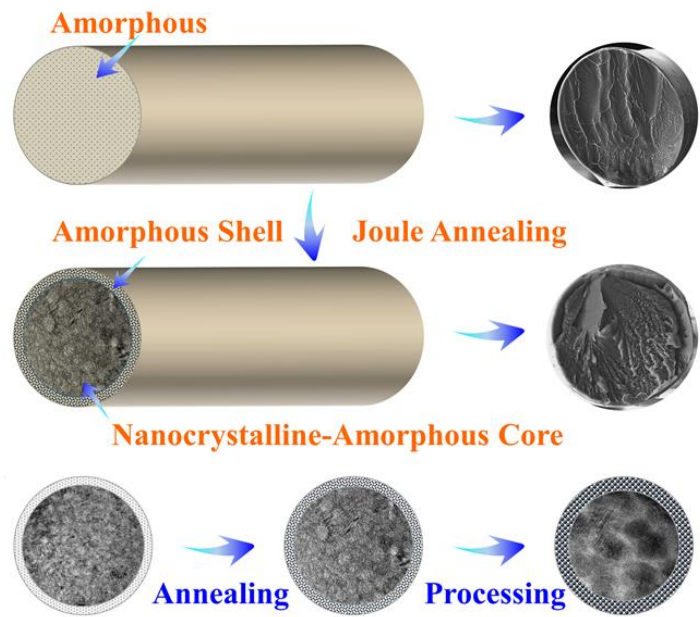